# The Coulomb Blockade Resonant Breakdown Caused By The Quantum Dot Mechanical Oscillations

A. G. Pogosov, M. V. Budantsev, A. A. Shevyrin,

A. E. Plotnikov, A. K. Bakarov, A. I. Toropov

*Institute of Semiconductor Physics, SB RAS, Novosibirsk, 630090, Russia*

E-mail: *pogosov@isp.nsc.ru*

Influence of forced mechanical vibrations of a suspended single-electron transistor on electron tunneling through the quantum dot limited by the Coulomb blockade is investigated. It is shown that mechanical oscillations of the quantum dot lead to the Coulomb blockade breakdown, shown in sharp resonant peaks in the transistor conductance dependence on the excitation frequency at values corresponding to the mechanical oscillations eigen modes. Physical mechanism of the observed effect is considered. It is presumably connected with oscillations of the mutual electrical capacitances between the quantum dot and surrounding electrodes.

PACS: 73.23.Hk, 73.63.Kv, 85.85.+j

The Coulomb blockade phenomenon [1] is usually investigated in single-electron transistors made lithographically in a rigid semiconductor bulk. Recently investigations of so-called nano-electromechanical systems [2] suspended over a substrate have begun. These structures are made of conducting semiconductor membranes (cantilevers) separated from the substrate by means of a selective etching of a sacrificial layer situated between the membrane and the substrate. Suspended single-electron transistor is one of the structures of this type that was also investigated [3-5]. These investigations have shown that these transistors have high charging energies due to the separation of the quantum dot from the substrate having high dielectric constant leading to essential decrease in the quantum dot capacitance [5].

Furthermore the separation of a transistor from a rigid substrate leads to appearance of additional mechanical degrees of freedom. As a result so-called elastic deformation blockade was found in a single-electron transistor formed in a suspended quantum wire. This blockade is additional to the Coulomb blockade and results from wire mechanical deformations, accompanying charge tunneling through the quantum dot [5, 6]. In the present paper we investigate the influence of forced mechanical oscillations on the electron transport in a suspended single-electron transistor.

The single-electron transistor was made from 110 nm thick AlGaAs/GaAs heterostructure with a two-dimensional electron gas in the 100 $\mathring{A}$ thick GaAs layer grown by means of molecular-beam epitaxy over a 400 nm thick AlAs sacrificial layer and a thick gallium-arsenide substrate. A lateral geometry was defined by electron lithography and subsequent anisotropic plasma etching. Then the sacrificial layer was selectively etched in a water solution of the hydrofluoric acid [4, 5].

The sample represents itself a 700 nm quantum dot, connected with a nanowire (width $W = 300$ nm, thickness $t = 110$ nm and length $l = 1\ \mu m$). The width of the nanowire was chosen to suppress its conductivity due to the depletion effect. The single-electron transistor is equipped with 5 side gates (see Fig. 1). Alternating RF voltage exciting mechanical vibrations of the nanowire was applied to the gate *G1*. The magnitude of RF voltage was chosen in the range from 0 to 100 mV, and the frequency from 100 kHz to 1 GHz. The gates *G2* and *G3* were used to govern the quantum dot electrochemical potential. The voltages applied to them were equal during the measurements $V_{G2} = V_{G3} = V_G$. The gates *G4* and *G5* were used to adjust the transparency of tunnel barriers. Corresponding voltages were chosen in a way to obtain the most pronounced Coulomb blockade peaks. The conductance was measured at ac signal by applying a voltage between the drain *D* and the source *S* oscillating with a frequency of 70 Hz and amplitude of 30 μV. All the measurements were carried out at the temperature 4.2 K.

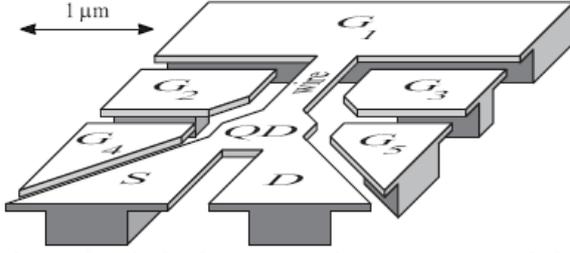

Fig. 1 The single-electron transistor geometry. Labels: G1..G5 – gates, S – source, D – drain, QD – quantum dot.

Fig. 2 shows the dependence of the single-electron transistor conductance on the dc gate and source-drain voltages $V_G$ and $V_{SD}$ at zero RF-field. One can see an area of low conductance near the point $V_G = -0.3\ V$, $V_{SD} = 0\ V$ corresponding to the Coulomb blockade regime (the center of the Coulomb blockade diamond). The charging energy determined as a half of the Coulomb blockade diamond size along the $V_{SD}$-axis is 3.5 meV or 40 K in temperature units.

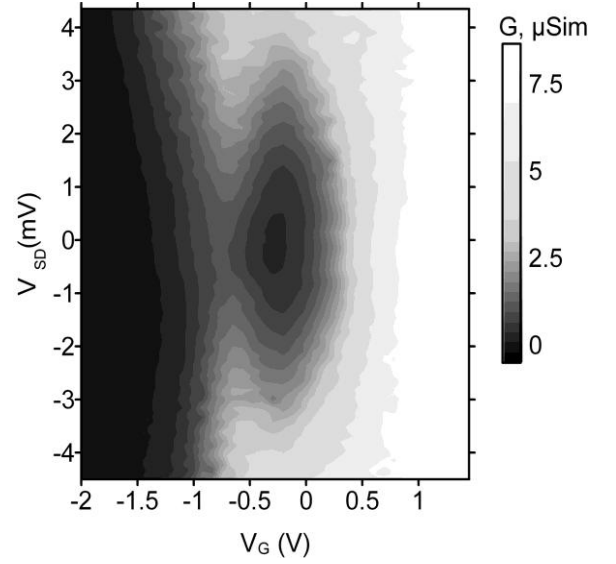

Fig. 2 The conductance of the single electron transistor as a function of the gate voltage $V_G$ (X-axis) and the source-drain voltage $V_{SD}$ (Y-axis).

The conductance dependence on the frequency of the ac voltage $V_{G1}$ of the fixed amplitude 10 mV for different sample states obtained in different immersions into liquid helium are presented in Fig. 3. Source-drain voltage is $V_{SD} = 0\ V$. Dc gate voltages are chosen in a way to turn the transistor into the Coulomb blockade regime in the absence of RF signal. These voltages are different for different sample states. This is related to the frozen potential of random impurities, which are known to occupy different charge states after temperature cycling.

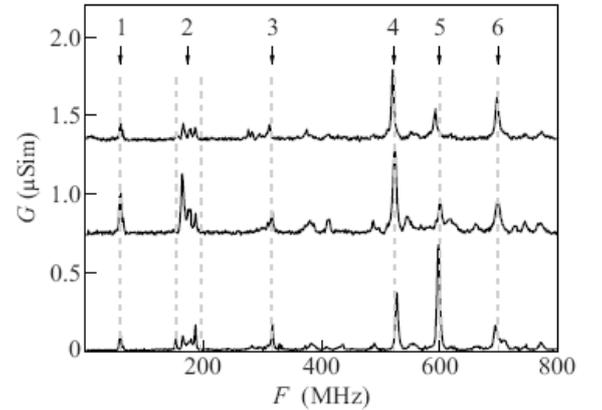

Fig. 3 The dependence of the conductance G of the single-electron transistor turned into the Coulomb blockade regime on the frequency F of the ac voltage with the amplitude 10 mV, applied to the gate G1. The experimental curves are obtained after different immersions of the sample into liquid helium. The peaks on the figure are denoted by numbers: 1 – 60 MHz, 2 – 150-190 MHz, 3 – 316 MHz, 4 – 526 MHz, 5 – 600 MHz, 6 – 700 MHz. Two upper curves are lifted along the Y-axis by 0.7 μSim and 1.3 μSim correspondingly.

Fig. 3 shows the resonant Coulomb blockade breakdown (positive spikes of the transistor conductance) at certain frequency values for all of the presented curves. It has been found that the peaks positions remain unchanged after temperature recycling. In Fig. 3 these peaks are denoted by the numbers: 1 – 60 MHz, 2 – frequency band 150-190 MHz, 3 – 316 MHz, 4 - 526 MHz, 5 - 600 MHz, 6 – 700 MHz.

The fact that the resonant frequencies remain unchanged after different temperature cycling means that the observed resonances are caused by the transistor mechanical vibrations, eigen frequencies of which are independent on electrostatics and obviously on the random impurity potential.

Estimations of eigen frequencies of different longitudinal and transversal nanowire vibrations were carried out. The gallium arsenide parameters used in the estimations were: Young modulus $E = 86$ GPa, density $\rho = 5390 \ g/m^3$.

It is known that the nanowire transversal vibrations can be divided into four types [7]: dilatational, torsional, shear and flexural. Dilatational vibrations correspond to either alternate contraction along one of the directions and widening along another one, or simultaneous contraction-widening along both directions. Torsional vibration is a periodic rotation of the nanowire cross section along the longitudinal axis remaining its shape undistorted [9]. Shear oscillations, in contrast, makes the nanowire cross-section take a form of a parallelogram with periodically oscillating angles.

Estimations show that only flexural mode frequencies fall into the measurement bandwidth (several hundreds MHz). Consider the nanowire as a pin-type rod [8]. Flexural waves in such a rod are described by the equation

$$\rho S \ddot{X} = E I_y \frac{\partial^4 X}{\partial z^4}$$

where the $z$-axis is directed along the rod axis, $X$ is a transversal displacement, $S = Wt$ is an area of the cross section and $I_y = Wt^3/12$ is the cross section mechanical momentum relative to the $y$ axis. Solving this equation with the boundary conditions corresponding to the fixed edges,

$$X(0) = X(l) = 0,$$

$$\frac{\partial X}{\partial z}(0) = \frac{\partial X}{\partial z}(l) = 0$$

one can find a transcendental algebraic equation

$$\cos(kl) \operatorname{ch}(kl) = 1, \qquad k^4 = \omega^2 \frac{\rho S}{E I_y}$$

Solving it one can find the eigen frequencies of the flexural vibrations defined by the formula

$$\nu_n = \alpha_n \sqrt{\frac{E}{\rho}} \frac{t}{l^2},$$

where $\alpha_1 = 1.027, \alpha_2 = 2.83, \alpha_3 = 3.56$ etc.

The length of the nanowire with a vibrating part of a quantum dot is about $l = 1.5$ μm, and its average width is about $W = 350$ nm. Corresponding flexural vibrations frequencies are $\nu_1^t = 200 \ MHz, \nu_2^t = 550 \ MHz, \nu_3^t = 700 \ MHz, \nu_1^W = 640 \ MHz$ where indexes t and W denote the vibrations along the thickness and along the width correspondingly. Estimated frequencies are in a good agreement with those observed in the experiment (see Fig. 3).

The resonance at the frequency 60 MHz (see peak 1 at Fig. 3) deserves separate consideration. In a selective etching process the sacrificial layer is removed both under the nanowire, and unavoidably under the source, drain and gates edges. We believe that the resonance at 60 MHz corresponds to vibrations of the quantum dot that is suspended on the membranes forming the source and the drain areas. The estimated value of the corresponding frequency can be obtained calculating the vibration frequency of the infinite undercut edge of the suspended membrane. Estimations of the frequencies are made in a way similar to the described above [8] with the boundary conditions

$$X(0) = 0, \qquad \frac{\partial X}{\partial z}(0) = 0$$

$$\frac{\partial^2 X}{\partial z^2}(l_{et}) = \frac{\partial^3 X}{\partial z^3}(l_{et}) = 0$$

for the membrane with one fixed edge and free another one. Here $l_{et}$ denotes the width of the undercut area of the membrane edge. The frequency 60 MHz corresponds to $l_{et} = 2.7\,\mu$. This value is in a good agreement with the experiment, since the actual width of undercut areas is of the order of 2 μ, and the quantum dot size 0.7 μ should be added to it.

The mechanical vibrations induced breakdown of the Coulomb blockade observed in our experiments could be explained in the following way. At zero source-drain voltage the single-electron transistor conductance is defined by the quantum dot electrochemical potential. The latter is defined by the gate voltage $V_G$ and the capacitances between each of the electrodes and the QD. At constant capacities the transistor can be switched from the closed state to the open one by changing the gate voltage $\Delta V_G$. At constant voltages on the electrodes this can be achieved by changing the capacitances. Transistor mechanical vibrations lead to just this change in the capacitances. The main change takes place in the QD-gate capacitance and the QD-substrate capacitance while the QD-source and QD-drain capacitances remain practically unchanged, because the source and the drain are rigidly connected to the QD and thus vibrate synchronously. For a rough estimate one can assume that the change in the gate capacitance leads to the same conductance change as the gate voltage bias $\frac{\Delta C_G}{C_G} = \frac{\Delta V_G}{V_G}$. For our experiment this corresponds to $\frac{\Delta C_G}{C_G} \approx 1$. This is quite possible for the quantum wire vibrating with almost the maximal amplitude limited by underlying substrate, which is quite reasonable for resonant vibrations.

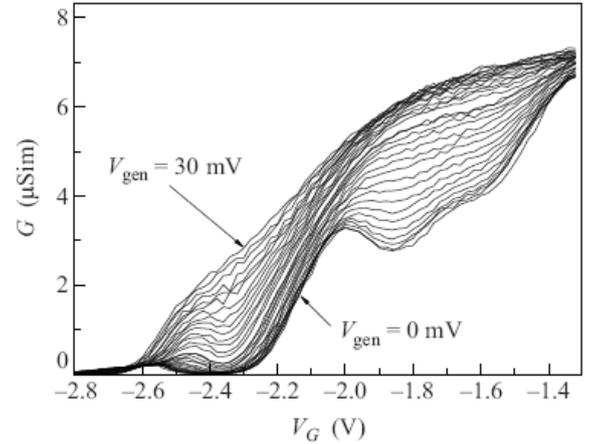

Fig. 4 A Series of G(V$_G$) curves measured at different amplitudes V$_{RF}$ of RF voltage, changing from 0 mV (the lowest curve) to 30 mV (the upper curve).

A series of the $G(V_G)$ curves measured at different applied RF – voltages is shown in Fig. 4. One can see that besides the discussed effect of the Coulomb blockade breakdown the applied RF-voltage of sufficient amplitude (more than 10 mV) leads to the shift of the Coulomb blockade peaks towards zero gate voltage. Such a behavior can be explained assuming that the quantum dot displacement during vibrations leads to a non-linear change in the QD-gate and QD-substrate capacitances. This means that the magnitudes of these capacitances averaged over the vibrations period change with the increase of the vibrations amplitude, which leads to the renormalization of the $V_G$ scale.

A high sensitivity of the conductance of a Coulomb-blockaded quantum dot to mechanical vibrations points to possible applications of single-electron transistors for measurements of small mechanical displacements. Usually, measurements of cantilevers small displacements are based on the detection of deviations of a laser beam reflected from it, which is inconvenient in many instances. A displacement detection based on the measurements of a cantilever-substrate impedance is reported in [10]. However these measurements require sensitive devices. The described measurements of the Coulomb blockade in the single-electron transistor placed on the cantilever are free from these disadvantages. Moreover, mechanical vibrations of the system investigated are characterized by a rather high quality factor *Q ≈ 100* and this is not the limit. Thus, the quality factor exceeding $10^4$ is reported in [10]. This offers a challenge to the mass spectrometry of the bio-molecules based on the single-electron transistor [11].

During the preparation of the present paper for publication we had known that relation between the Coulomb blockade and mechanical vibrations is investigated in carbon nanotubes used as a vibrating nanowire [12].